\documentclass[11pt]{article}

\usepackage[margin=2.6cm]{geometry}
\usepackage{amsmath,amssymb}
\usepackage{graphicx}
\usepackage[font=small,labelfont=bf]{caption}

\newcommand{\Om}{\Omega_{m}}
\newcommand{\OL}{\Omega_{\Lambda}}
\newcommand{\OPhi}{\Omega_{\Phi}}
\newcommand{\weff}{\omega_{\mathrm{eff}}}

\title{\textbf{No late-time role for adiabatic torsion:\\
a no-go result for Hubble-cutoff\\ holographic dark energy in Einstein--Cartan cosmology}}

\author{
Fernando Izaurieta\footnote{\texttt{fernando.izaurieta@uss.cl}}\\
{\small Universidad San Sebasti\'an, Lientur 1457, Concepci\'on, Chile}
\and
Samuel Lepe\footnote{\texttt{samuel.lepe@pucv.cl}}\\
{\small Instituto de F\'isica, Pontificia Universidad Cat\'olica de Valpara\'iso,}\\
{\small Av.\ Brasil 2950, Valpara\'iso, Chile}
\and
Cristian Quinzacara\footnote{\texttt{cristian.quinzacara@uss.cl}}\\
{\small Universidad San Sebasti\'an, Lientur 1457, Concepci\'on, Chile}
}
\date{July 2026}

\begin{document}
\maketitle

\begin{abstract}
\noindent
In Friedmann cosmology with Einstein--Cartan torsion, the homogeneous torsion mode compatible with a separately conserved matter sector scales as $\Phi\propto a^{-3}$ and enters the Friedmann constraint as a stiff component of negative energy density, $-3\Phi^{2}\propto a^{-6}$. It has recently been claimed that this mode rescues the Hubble radius as an infrared cutoff for holographic dark energy, producing late-time acceleration and a phantom-divide crossing of possible relevance to DESI. We show that it cannot. With the Hubble cutoff the holographic density drops out of the deceleration parameter, and the only accelerating regime is the transient window $\bar{a}\leq a<4^{1/3}\bar{a}$ around the torsion bounce at $H(\bar{a})=0$; placing that window at observable redshifts would force the Hubble rate to vanish in our recent past, against the measured expansion history. Requiring a viable history yields nested upper bounds on $\OPhi\equiv(\Phi_{0}/H_{0})^{2}$: fitting the official DESI DR2 BAO likelihood gives $\OPhi<8.7\times10^{-4}$ ($95\%$ CL), the existence of a spectroscopically confirmed galaxy at $z=14.32$ gives $8.4\times10^{-5}$, the CMB gives $3.1\times10^{-10}$, and Big Bang nucleosynthesis gives $5\times10^{-24}$. Today's imprint on the dark energy equation of state, $|1+\omega_{0}|\leq2\OPhi/\OL$, falls short of the DESI preference by two orders of magnitude at best and twenty-two at worst, and on the phantom side. In the Granda--Oliveros cutoff, torsion deepens rather than prevents the big-rip singularity. An appendix derives the Friedmann pair from the Einstein--Cartan field equations and shows that the $a^{-3}$ scaling is the kinematics of a diluting spin fluid; escaping the no-go requires breaking exactly that.
\end{abstract}

\section{Introduction}
\label{sec:intro}

The DESI DR2 baryon acoustic oscillation (BAO) measurements, in combination with supernovae and the CMB, prefer an evolving dark energy over a cosmological constant at the $2.8$--$4.2\sigma$ level~\cite{DESI-DR2}. Whatever the fate of that preference, it has reopened a question that refuses to stay closed: which geometric degrees of freedom, if any, could imitate a dynamical dark energy at low redshift? Spacetime torsion is a recurring candidate. It is not an exotic addition but the part of the affine connection that general relativity sets to zero by hand; in Einstein--Cartan (EC) gravity it is sourced by intrinsic spin and switches on wherever fermions are dense enough~\cite{Hehl1976,Trautman1973,Poplawski2012,Brechet2008,KTBI2019}. Torsion cosmologies have accordingly been confronted with data more than once, as substitutes for dark matter~\cite{TS2011,Pereira2022} and, recently, against the DESI era itself~\cite{Liu2025}.

A proposal by Yun and Lee~\cite{YunLee2026,YunLee2024} puts torsion to work on an old problem of holographic dark energy (HDE). In HDE the dark energy density is tied to an infrared cutoff $L$, $\rho_{hol}=3c^{2}L^{-2}$ (in units $8\pi G=1$; $c^{2}<1$ is Li's dimensionless parameter\footnote{An unfortunate but by now canonical notation clash: $c$ here is Li's constant, not the speed of light, which is set to one throughout.}). The natural choice $L=H^{-1}$, the Hubble radius, famously fails: the holographic density then tracks the dominant component and cannot accelerate the expansion, which is why the field moved to the future event horizon and its causality troubles~\cite{Li2004,Hsu2004}. Yun and Lee claim that in EC gravity the scalar torsion mode $\Phi$ changes this verdict: the deceleration parameter is shifted toward negative values, the effective equation of state can cross the phantom divide, and the Hubble cutoff becomes viable after all, with possible relevance to the DESI anomaly~\cite{YunLee2026}.

In this paper we show that the claim fails, and that it fails for reasons reaching well beyond this particular model. The mode in question is fixed by self-consistency: if matter is separately conserved, the torsion scalar obeys $\dot\Phi+3H\Phi=0$, hence $\Phi\propto a^{-3}$, and it enters the Friedmann constraint as $-3\Phi^{2}\propto a^{-6}$, a \emph{negative} energy density with stiff equation of state. Appendix~\ref{app:derivation} traces both statements back to the EC field equations and to the kinematics of spin fluids.

The first casualty is the holographic component itself. Being proportional to $H^{2}$, it drops out of the deceleration parameter identically: $q$ is the same function of $(\rho,\Phi)$ with or without it. Whatever acceleration the model produces belongs to torsion alone, and torsion produces it only in the aftermath of its bounce. The constraint $H^{2}\geq0$ enforces a turning point $H(\bar{a})=0$, the classical EC bounce~\cite{Trautman1973,Poplawski2012,Brechet2008}, and confines the entire accelerating regime to $\bar{a}\leq a<4^{1/3}\bar{a}$, a window less than a factor $1.6$ wide in scale factor. The window through which torsion could accelerate the universe closes before any observer can look through it: dragging it to redshifts where acceleration is actually observed forces $H$ to vanish in our recent past, and the measured expansion history says otherwise.

What turns this observation into numbers is a requirement almost embarrassing in its modesty: the universe must contain its own past.

That requirement yields nested upper bounds on $\OPhi\equiv(\Phi_{0}/H_{0})^{2}$. A fit to the DESI DR2 BAO distances gives $\OPhi<8.7\times10^{-4}$ at $95\%$ confidence. The existence of the $z=14.32$ galaxy JADES-GS-z14-0~\cite{Carniani2024} tightens this to $8.4\times10^{-5}$; the CMB, to $3.1\times10^{-10}$; Big Bang nucleosynthesis (BBN), to $5\times10^{-24}$. Each rung rests on strictly weaker assumptions than the one before, and the last one settles the question. The imprint left on the dark energy equation of state today, $|1+\omega_{0}|\leq2\OPhi/\OL$, sits at least two orders of magnitude below the DESI preference (twenty-two, once BBN is enforced), and on the wrong side of $-1$ besides.

Our target, we stress, is narrow: one scaling, $\Phi\propto a^{-3}$, and the single assumption that forces it. Torsion cosmology at large survives this paper, and Sec.~\ref{sec:discussion} spells out what it would have to give up to leave a late-time imprint. Throughout, $8\pi G=c_{\rm light}=1$, $a_{0}=1$ today, and dots denote cosmic-time derivatives.

\section{The adiabatic torsion mode and its bounce}\label{sec:mode}

Our starting point is the pair of modified Friedmann equations obtained in~\cite{YunLee2026} (Eqs.~(3.28)--(3.29) of the arXiv v1) for a flat Friedmann--Lema\^itre--Robertson--Walker metric in EC gravity with a homogeneous scalar torsion mode $\Phi(t)$,
\begin{align}
3H^{2} &= \rho-3\Phi^{2},\label{eq:fried1}\\
\dot{H}+H^{2} &= -\frac{1}{6}\left(\rho+3p-12\Phi^{2}\right).\label{eq:fried2}
\end{align}
Both reduce to general relativity for $\Phi=0$; their EC pedigree is reviewed in Appendix~\ref{app:derivation}. Combining them gives an identity we will lean on more than once,
\begin{equation}
\dot{\rho}+3H\left(\rho+p\right)=6\Phi\left(\dot{\Phi}+3H\Phi\right),\label{eq:bianchi}
\end{equation}
which says that matter and torsion may exchange energy. If one demands that the matter sector be \emph{separately} conserved (equivalently, that the expansion be adiabatic, $TdS=0$, for the fluid alone\footnote{Using $TdS=d(\rho V)+p\,dV$ with $V\propto a^{3}$, the left-hand side of \eqref{eq:bianchi} is $T\dot{S}/V$. Adiabaticity is an assumption; what happens when it is dropped is the subject of Sec.~\ref{sec:discussion}, and Appendix~\ref{app:derivation} explains why a diluting spin fluid enforces it automatically.}), then
\begin{equation}
\dot{\Phi}+3H\Phi=0\qquad\Longrightarrow\qquad\Phi(a)=\Phi_{0}\,a^{-3}.\label{eq:phia}
\end{equation}
This is the scaling found in~\cite{YunLee2026} and the one we examine throughout. The torsion contribution to~\eqref{eq:fried1}--\eqref{eq:fried2} can be bookkept as an effective fluid with
\begin{equation}
\rho_{\Phi}=-3\Phi_{0}^{2}\,a^{-6},\qquad p_{\Phi}=\rho_{\Phi},\qquad \omega_{\Phi}=+1 , \label{eq:stiff}
\end{equation}
a stiff component of \emph{negative} energy density. The combination is an old acquaintance: it is how the spin--spin contact interaction of a Weyssenhoff fluid enters EC cosmology, where it powers the nonsingular bounce at high density~\cite{Trautman1973,Brechet2008,Poplawski2012} (Appendix~\ref{app:derivation}). What follows is therefore the physics of the classical EC bounce, examined at the opposite end of the expansion history. The question is how much room the data leave for it there.

\subsection{The turning point}

With a single barotropic fluid $p=\omega\rho$, $\rho=\rho_{0}a^{-3(1+\omega)}$, the constraint \eqref{eq:fried1} enforces $\rho\geq3\Phi^{2}$, i.e.
\begin{equation}
a\geq\bar{a}\equiv\Theta^{\frac{1}{3(1-\omega)}}, \qquad \Theta\equiv\frac{3\Phi_{0}^{2}}{\rho_{0}} , \label{eq:abar}
\end{equation}
for $\omega<1$. At $a=\bar{a}$ the Hubble rate vanishes while, for dust, $\dot{H}(\bar{a})=\tfrac{3}{2}\Phi^{2}(\bar{a})>0$: this is a bounce, not an end. The universe contracts, reaches $\bar{a}$, and re-expands. In the early-universe setting this is the celebrated singularity avoidance of EC gravity; here it reappears as the boundary of the ``story'' available to any late-time torsion scenario.

\subsection{Effective equation of state and the two windows}

From \eqref{eq:fried1}--\eqref{eq:fried2}, the deceleration parameter $q=-1-\dot{H}/H^{2}$ reads
\begin{equation}
q=\frac{\rho+3p-12\Phi^{2}}{2\left(\rho-3\Phi^{2}\right)}=\frac{1}{2}\left(1+3\,\weff\right), \qquad \weff=\frac{\omega\rho-3\Phi^{2}}{\rho-3\Phi^{2}} . \label{eq:qweff}
\end{equation}
For dust ($\omega=0$), $\weff=-3\Phi^{2}/(\rho-3\Phi^{2})$, and the classification is immediate:
\begin{align}
\text{phantom }(\weff<-1):&\qquad \bar{a}<a<2^{1/3}\bar{a}, \label{eq:phantomwin}\\[2pt]
\text{quintessence }(-1<\weff<-\tfrac{1}{3}):&\qquad 2^{1/3}\bar{a}<a<4^{1/3}\bar{a}, \label{eq:quintwin}\\[2pt]
\text{acceleration }(q<0):&\qquad \bar{a}\leq a<4^{1/3}\bar{a}. \label{eq:accwin}
\end{align}
The universe exits the bounce super-accelerating, crosses the phantom divide at $a=2^{1/3}\bar{a}$, stops accelerating at $a=4^{1/3}\bar{a}$, and tends to $q\to\tfrac{1}{2}$, $\weff\to0$ thereafter: cold matter with a rapidly fading memory of the bounce.\footnote{The divergence of $\weff$ at $a\to\bar{a}$ is the standard effective-fluid artifact at a bounce ($H\to0$ with $\dot H>0$); every bouncing model produces it.} The entire ``dark energy mimicry'' of this torsion mode spans a factor $4^{1/3}\simeq1.59$ in scale factor, anchored to the bounce. That anchor is the whole problem.

\section{Why the Hubble cutoff is not rescued}\label{sec:holographic}

Pairing holographic dark energy with torsion has some history already~\cite{YunLee2024,LiChen2023}; the Hubble-radius version of~\cite{YunLee2026} is its sharpest form, and the one with the most to gain, since the Hubble cutoff is the one choice that needs rescuing.

\subsection{The holographic component is dynamically inert}

Let us begin with Li's density, $\rho_{hol}=3c^{2}H^{2}$ with $c^{2}<1$~\cite{Li2004,Hsu2004}. Note first that a universe containing \emph{only} $\rho_{hol}$ and torsion is impossible: Eq.~\eqref{eq:fried1} would demand $(1-c^{2})H^{2}=-\Phi^{2}<0$. Matter is mandatory. With dust $\rho$ (separately conserved) alongside, the constraint becomes
\begin{equation}
3\left(1-c^{2}\right)H^{2}=\rho-3\Phi^{2}, \label{eq:hde-constraint}
\end{equation}
and differentiating with $\dot{\rho}=-3H\rho$ and $\dot{\Phi}=-3H\Phi$, it is straightforward to obtain
\begin{equation}
q=\frac{1}{2}\left(1-\frac{9\Phi^{2}}{\rho-3\Phi^{2}}\right)=\frac{1}{2}\left[1-\frac{3}{1-c^{2}}\left(\frac{\Phi}{H}\right)^{2}\right]=\frac{\rho-12\Phi^{2}}{2\left(\rho-3\Phi^{2}\right)} . \label{eq:qhde}
\end{equation}
Compare \eqref{eq:qhde} with \eqref{eq:qweff} at $\omega=0$: \emph{the holographic component has dropped out}. The deceleration parameter is the same function of $(\rho,\Phi)$ with or without $\rho_{hol}$; a density proportional to $H^{2}$ merely renormalizes the constraint by $(1-c^{2})$ and inherits whatever dynamics the rest dictates. This is the EC version of the tracker pathology that has plagued the Hubble cutoff since the beginning~\cite{Hsu2004,Li2004}, and it survives torsion untouched.

Acceleration ($q<0$) therefore occurs if and only if $\rho<12\Phi^{2}$, which by~\eqref{eq:abar} is again the window \eqref{eq:accwin}: $a<4^{1/3}\bar{a}$. There \emph{is} acceleration in this model, as claimed in~\cite{YunLee2026}, and it can even occur at ``weak torsion,'' $(\Phi/H)^{2}>(1-c^{2})/3$. But it is the bounce transient of Sec.~\ref{sec:mode} wearing holographic clothes, and it lives where it always lived. At late times $(\Phi/H)^{2}\to0$ and $q\to\tfrac{1}{2}$: cold dark matter in disguise, with no acceleration and no phantom crossing at any observable epoch. The claim of~\cite{YunLee2026} conflates an instantaneous parameter choice with an expansion history. One may certainly pick $(\Phi/H)^{2}$ large enough to make $q<0$ \emph{now}; Eq.~\eqref{eq:qhde} then forces ``now'' to sit within a factor $1.6$ in scale factor of $H=0$.

\subsection{Granda--Oliveros cutoff: torsion deepens the rip}

Consider now the Granda--Oliveros density, $\rho_{hol}=3(\alpha H^{2}+\beta\dot{H})$~\cite{GO2008}. We work in its natural habitat, the dark-energy dominated late universe, and neglect matter (its inclusion only postpones the asymptotics below); this is the regime in which the big-rip behaviour of this cutoff arises~\cite{Lepe2026}. The constraint~\eqref{eq:fried1} gives
\begin{equation}
\dot{H}=\frac{1}{\beta}\left[\left(1-\alpha\right)H^{2}+\Phi^{2}\right] \qquad\Longrightarrow\qquad q=-1-\frac{1}{\beta}\left[\left(1-\alpha\right) +\left(\frac{\Phi}{H}\right)^{2}\right]. \label{eq:GOq}
\end{equation}
For the usual parameter region $\alpha<1$, $\beta>0$, Eq.~\eqref{eq:GOq} gives $q<-1$ \emph{identically}: the torsion term adds to the super-acceleration instead of moderating it. In the regime $\Phi^{2}\ll\rho_{hol}$ the solution is the big-rip trajectory
\begin{equation}
H(t)=\frac{\beta}{1-\alpha}\,\frac{1}{t_{s}-t},
\qquad
t_{s}=t_{0}+\frac{\beta}{1-\alpha}H_{0}^{-1}, \label{eq:bigrip}
\end{equation}
as in the torsionless case~\cite{Lepe2026}; and since $\Phi^{2}\propto a^{-6}$ dies while $H$ grows, torsion becomes ever more irrelevant on approach to $t_{s}$. Torsion does not regularize the future singularity of this cutoff; if anything, it hastens it. This complements the phantom asymptotics studied in~\cite{Lepe2022,Lepe2023}.

\section{The no-go from the expansion history}\label{sec:nogo}

\begin{figure}[t]
\centering
\includegraphics[width=0.62\textwidth]{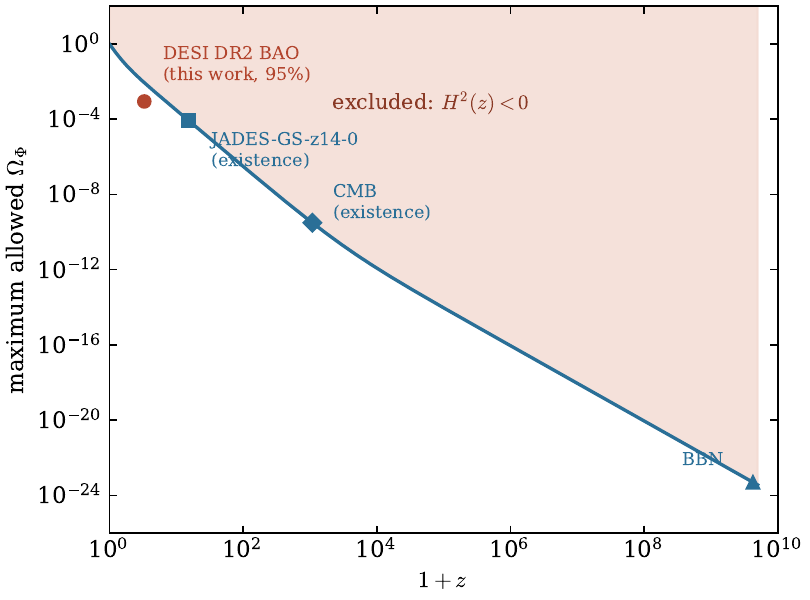}
\caption{The closing window. The curve is the maximum $\OPhi$ compatible with $H^{2}(z)>0$ at each redshift, Eq.~\eqref{eq:ladder}; the shaded region above it leaves no universe at that epoch. Markers show the ladder of Table~\ref{tab:ladder}: the DESI DR2 MCMC bound (Sec.~\ref{sec:mcmc}), the existence bounds from JADES-GS-z14-0 and the CMB, and the BBN bound, Eq.~\eqref{eq:bbnbound}.} \label{fig:window}
\end{figure}

We now embed the torsion mode in the realistic budget. Define
\begin{equation}
\OPhi\equiv\left(\frac{\Phi_{0}}{H_{0}}\right)^{2}\geq0 ,
\qquad
E^{2}(z)\equiv\frac{H^{2}}{H_{0}^{2}}=\Om\left(1+z\right)^{3}+\Omega_{r}\left(1+z\right)^{4}+\OL-\OPhi\left(1+z\right)^{6}, \label{eq:E2}
\end{equation}
with flatness fixing $\OL=1-\Om-\Omega_{r}+\OPhi$.\footnote{Ref.~\cite{YunLee2026} absorbs the sign into the density parameter, defining $\Omega_{\Phi_{0}}=-(\Phi_{0}/H_{0})^{2}\leq0$. We keep $\OPhi\geq0$ and display the sign explicitly in Eq.~\eqref{eq:E2}, so that the constraints below read as ordinary upper limits.} The torsion term is the fastest-growing one toward the past. Requiring $H^{2}(z)>0$ up to some redshift $z_{\max}$ that the universe demonstrably reached is the mildest of all cosmological priors: it asks only that the universe contain its own past.\footnote{Nothing is fitted here. An object observed at $z_{\max}$ exists, therefore the expansion history extends at least that far, and $\bar{a}<(1+z_{\max})^{-1}$.} Neglecting radiation for a moment, $H^{2}$ vanishes at
\begin{equation}
\left(1+z_{*}\right)^{3}=\frac{\Om+\sqrt{\Om^{2}+4\,\OPhi\OL}}{2\,\OPhi} \;\simeq\;\frac{\Om}{\OPhi}
\quad\text{for }\OPhi\ll\Om^{2} . \label{eq:zstar}
\end{equation}
The numbers are unforgiving. For torsion to contribute even $1\%$ of today's critical density, $\OPhi=10^{-2}$, Eq.~\eqref{eq:zstar} puts the turning point at $z_{*}\simeq2.2$: the universe would have no history beyond the redshift at which DESI measures the Lyman-$\alpha$ BAO ($z_{\rm eff}=2.33$)~\cite{DESI-DR2}. For $\OPhi=10^{-3}$, $z_{*}\simeq5.7$, excluded by every quasar and galaxy beyond $z=6$. Each rung of observed history pushes the bound down as $(1+z_{\max})^{-6}$ against the growth of everything else:
\begin{equation}
\OPhi\;\leq\; \frac{\Om\left(1+z_{\max}\right)^{3} +\Omega_{r}\left(1+z_{\max}\right)^{4}+\OL} {\left(1+z_{\max}\right)^{6}} . \label{eq:ladder}
\end{equation}
Table~\ref{tab:ladder} and Fig.~\ref{fig:window} give the resulting ladder, using $\Om=0.3$, $\Omega_{r}=9.1\times10^{-5}$: the spectroscopically confirmed galaxy JADES-GS-z14-0 at $z=14.32$~\cite{Carniani2024} gives $\OPhi<8.4\times10^{-5}$; the CMB at $z\simeq1090$~\cite{Planck2018} gives $\OPhi<3.1\times10^{-10}$; and BBN at $T\simeq1\,$MeV, i.e. $1+z\simeq4.3\times10^{9}$,\footnote{We ignore the $e^{\pm}$ reheating of the photon bath; at twenty-four orders of magnitude, factors of order unity are a luxury we can afford.} gives
\begin{equation}
\OPhi\;<\;\Omega_{r}\left(1+z_{\rm BBN}\right)^{-2}\simeq5\times10^{-24}. \label{eq:bbnbound}
\end{equation}
Demanding further that the torsion term not perturb the BBN expansion rate by more than the usual $\sim10\%$ budget~\cite{Fields2020} sharpens~\eqref{eq:bbnbound} by another order of magnitude. Either way, the conclusion does not move: \emph{any} adiabatic torsion mode consistent with the mere existence of the early universe is bounded twenty or more orders of magnitude below cosmological relevance today.

\begin{table}[t]
\centering
\caption{The ladder of upper bounds on $\OPhi=(\Phi_{0}/H_{0})^{2}$. Each row uses only the requirement stated; rows are independent and progressively stronger. The MCMC bound (Sec.~\ref{sec:mcmc}) marginalizes over $(\Om, H_{0}r_{d})$; the ``existence'' bounds use Eq.~\eqref{eq:ladder} with $\Om=0.3$, $\Omega_{r}=9.1\times10^{-5}$.} \label{tab:ladder}
\vspace{4pt}
\begin{tabular}{lll}
\hline\hline
Probe & Requirement & Bound on $\OPhi$ \\
\hline
DESI DR2 BAO (Sec.~\ref{sec:mcmc}) & fit, $95\%$ CL & $8.7\times10^{-4}$ \\
JADES-GS-z14-0, $z=14.32$ & existence & $8.4\times10^{-5}$ \\
CMB, $z\simeq1090$ & existence & $3.1\times10^{-10}$ \\
BBN, $1+z\simeq4.3\times10^{9}$ & $H^{2}>0$ & $5.0\times10^{-24}$ \\
BBN expansion-rate budget & $\lesssim10\%$ of $\rho_{r}$ & $\sim5\times10^{-25}$ \\
\hline\hline
\end{tabular}
\end{table}

Two translations make the irrelevance concrete. The torsion fluid~\eqref{eq:stiff} contributes $\Delta q_{0}=-2\OPhi$ to the deceleration parameter and shifts the effective dark energy equation of state to
\begin{equation}
\omega_{0}=-\,\frac{\OL+\OPhi}{\OL-\OPhi}\simeq-1-\frac{2\,\OPhi}{\OL} . \label{eq:w0}
\end{equation}
Even the weakest rung of the ladder (the BAO fit alone) caps the late-time imprint at $|\Delta q_{0}|<1.7\times10^{-3}$ and $|1+\omega_{0}|<2.5\times10^{-3}$, on the phantom side of $-1$; the BBN rung pushes both to the $10^{-23}$ level. The DESI DR2 preference for evolving dark energy asks for $|1+\omega_{0}|$ of order $0.25$ \emph{on the quintessence side}~\cite{DESI-DR2}: two orders of magnitude away at best, twenty-two at worst, and with the wrong sign throughout.

\section{Constraints from DESI DR2 BAO}
\label{sec:mcmc}

The bounds above assume nothing about the data beyond existence. To close the circle we let the most model-agnostic late-time dataset speak. We use the DESI DR2 BAO data vector and covariance exactly as released with the official collaboration likelihood~\cite{DESI-DR2}: thirteen entries, $D_{V}/r_{d}$ for the BGS sample plus the $(D_{M}/r_{d},\,D_{H}/r_{d})$ pairs of the six anisotropic tracers, with $r_{d}$ the drag-epoch sound horizon. Table~\ref{tab:bao} lists the corresponding means, uncertainties and correlations. We fit the flat model \eqref{eq:E2} (radiation neglected: it contributes less than $10^{-3}$ of $E^{2}$ over the fitted range, well below the data precision) with parameters $(\Om,\,\OPhi,\,H_{0}r_{d})$ and uniform priors $\Om\in(0.05,0.7)$, $\OPhi\in[0,10^{-2})$, $H_{0}r_{d}\in(7000,13000)\,$km\,s$^{-1}$, sampling with the affine-invariant ensemble sampler \texttt{emcee}~\cite{emcee} (40 walkers, $4000$ steps, first $1000$ discarded; acceptance $0.62$).\footnote{We restrict the fit to BAO deliberately: supernovae or CMB distance priors would only strengthen a bound that BBN, in any case, dwarfs.}

\begin{figure}[t]
\centering
\includegraphics[width=0.98\textwidth]{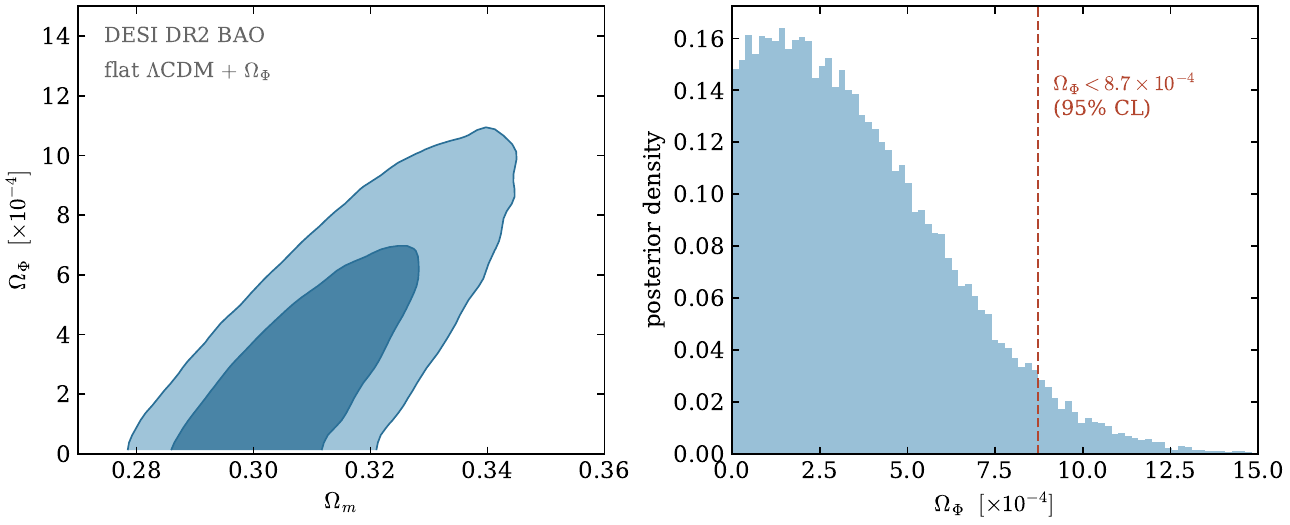}
\caption{Posterior from the DESI DR2 BAO fit of the flat $\Lambda$CDM$+\OPhi$ model. \emph{Left:} $68\%$ and $95\%$ credible regions in the $(\Om,\OPhi)$ plane. \emph{Right:} marginalized posterior of $\OPhi$ with the one-sided $95\%$ bound of Eq.~\eqref{eq:mcmcresult}.} \label{fig:posterior}
\end{figure}

As a validation, the $\OPhi=0$ run reproduces the published DESI DR2 $\Lambda$CDM fit: we find $\Om=0.2976\pm0.0087$ and $H_{0}r_{d}=(101.54\pm0.74)\times10^{2}\,$km\,s$^{-1}$, against $\Om=0.2975\pm0.0086$, $H_{0}r_{d}=(101.54\pm0.73)\times10^{2}\,$km\,s$^{-1}$ of~\cite{DESI-DR2}. Opening $\OPhi$ we obtain
\begin{equation}
\boxed{\;\OPhi<8.7\times10^{-4}\quad(95\%\ \mathrm{CL}),\;}
\qquad
\Om=0.312\pm0.013,
\qquad
H_{0}r_{d}=(100.8\pm0.9)\times10^{2}\,\mathrm{km\,s^{-1}}, \label{eq:mcmcresult}
\end{equation}
with $\Delta\chi^{2}_{\min}=\chi^{2}_{\Lambda\mathrm{CDM}}-\chi^{2}_{\Lambda\mathrm{CDM}+\Phi}=0.17$: the data express no preference whatsoever for torsion. The posterior is shown in Fig.~\ref{fig:posterior}. The mild upward drift of $\Om$ reflects the partial degeneracy between $-\OPhi(1+z)^{6}$ and the matter term over the BAO lever arm.

Even this weakest rung polices itself. Saturating the bound~\eqref{eq:mcmcresult} would put the turning point \eqref{eq:zstar} at $z_{*}\simeq6.1$, leaving no universe for the objects we observe beyond that redshift; the BAO data run out of constraining power just where the existence bounds take over.

\begin{table}[t]
\centering
\caption{DESI DR2 BAO measurements used in the fit, as distributed with the official collaboration likelihood~\cite{DESI-DR2}; uncertainties and the correlation coefficient $r$ between $D_{M}/r_{d}$ and $D_{H}/r_{d}$ are derived from the released covariance matrix.}
\label{tab:bao}
\vspace{4pt}
\begin{tabular}{lcccc}
\hline\hline
Tracer & $z_{\rm eff}$ & $D_{V}/r_{d}$ & $D_{M}/r_{d}$, $D_{H}/r_{d}$ & $r$ \\
\hline
BGS  & 0.295 & $7.942\pm0.076$ & --- & --- \\
LRG1 & 0.510 & --- & $13.588\pm0.168$, \ $21.863\pm0.429$ & $-0.452$ \\
LRG2 & 0.706 & --- & $17.351\pm0.180$, \ $19.455\pm0.334$ & $-0.395$ \\
LRG3+ELG1 & 0.934 & --- & $21.576\pm0.162$, \ $17.641\pm0.201$ & $-0.347$ \\
ELG2 & 1.321 & --- & $27.601\pm0.325$, \ $14.176\pm0.225$ & $-0.398$ \\
QSO  & 1.484 & --- & $30.512\pm0.764$, \ $12.817\pm0.518$ & $-0.494$ \\
Ly$\alpha$ & 2.330 & --- & $38.989\pm0.532$, \ $8.632\pm0.101$ & $-0.431$ \\
\hline\hline
\end{tabular}
\end{table}

\section{Discussion: what it would take}
\label{sec:discussion}

Every no-go is a list of assumptions read backwards, and ours has a short list. The load-bearing assumption is adiabaticity. Everything hinges on $\Phi\propto a^{-3}$, which follows from separate conservation through~\eqref{eq:bianchi}, and which Appendix~\ref{app:derivation} shows to be nothing deeper than dilution: for a comoving spin fluid, particle-number conservation gives $n\propto a^{-3}$, and the spin density follows suit. Evading it therefore costs real physics. One must let matter and torsion exchange energy, turning \eqref{eq:bianchi} into the definition of an interacting dark sector with $6\Phi(\dot\Phi+3H\Phi)$ as the coupling. A torsion history decaying more slowly than $a^{-3}$ (ultimately $\Phi\simeq$ const, mimicking $\omega\simeq-1$) then becomes kinematically available, and demands a spin density that refuses to dilute: progressive alignment, condensation, or some other mechanism operating at late times. In EC gravity torsion is algebraic and cannot outlive its source~\cite{Hehl1976}, so the mechanism must live in the matter sector explicitly. A fermionic dark matter component with evolving spin alignment is the natural candidate; its cosmology, including the effect on the $H_{0}$ tension, was explored in~\cite{ILV2020}, and we intend to return to it in this context.

Kinematically different torsion histories also escape, by construction. The steady-state mode $\phi\propto H$ of~\cite{KTBI2019} couples linearly to the expansion, modifies the matter scaling itself, and has been confronted with DESI DR2, supernovae and CMB data in~\cite{Liu2025}, with results consistent with zero at sub-percent precision. Dynamical torsion is a different animal altogether: in Poincar\'e gauge cosmology the propagating torsion modes can oscillate and drive late-time acceleration~\cite{SNY2008}, and nothing in our argument touches them. With~\cite{Liu2025} and the present work, the two simplest one-parameter torsion cosmologies now stand quantitatively constrained; the dynamical family remains open, on its own terms.

We have also stayed at the level of the background. Perturbations would only sharpen the verdict: a negative-density component growing as $a^{-6}$ into the past is even harder to hide in structure formation than in distances.

What remains of the Hubble cutoff itself? Its EC incarnation inherits, intact, the defect Hsu identified two decades ago~\cite{Hsu2004}: proportionality to $H^{2}$ makes it a spectator. Torsion adds a bounce, and the bounce brings a brief phantom display, but every part of it stays locked behind $z_{*}$. The window through which torsion could have accelerated the universe sits, and always sat, on the far side of that wall.

\section{Conclusions}
\label{sec:conclusions}

We have examined the recent claim~\cite{YunLee2026} that the adiabatic torsion mode of Einstein--Cartan cosmology rescues the Hubble radius as an infrared cutoff for holographic dark energy. It does not. The holographic density cancels out of the deceleration parameter, so the acceleration and the phantom crossing found in~\cite{YunLee2026} belong to the EC bounce transient, which the constraint $H^{2}\geq0$ confines to $\bar a\leq a<4^{1/3}\bar a$. A viable expansion history then bounds the mode's present abundance: $\OPhi<8.7\times10^{-4}$ from the DESI DR2 BAO distances ($95\%$ CL), $8.4\times10^{-5}$ from the existence of a $z=14.32$ galaxy, $3.1\times10^{-10}$ from the CMB, and $5\times10^{-24}$ from BBN. The corresponding shift in the dark energy equation of state today, $|1+\omega_{0}|\leq2\OPhi/\OL$, lies between two and twenty-two orders of magnitude below the DESI preference for evolving dark energy, and on the phantom side of $-1$, where DESI does not want it. In the Granda--Oliveros cutoff, torsion deepens rather than prevents the big rip.

The adiabatic mode, in short, is early-universe physics: it does its one spectacular thing near the bounce and afterwards dilutes faster than anything else in the inventory. If torsion has something to say about dark energy, it will have to say it where adiabaticity breaks, and that is where we plan to look next.

\subsection*{Acknowledgments}
F.~Izaurieta acknowledges support from ANID-FONDECYT Regular grant 1262414, and S.~Lepe acknowledges support from ANID-FONDECYT grant No.~1250969, from the Government of Chile. F.~Izaurieta is also grateful for the hospitality and stimulating environment of the GeomGravX Tartu Conference, where part of this work was written.

\subsection*{Data availability}
The BAO data used in Sec.~\ref{sec:mcmc} are the official DESI DR2 likelihood products~\cite{DESI-DR2}, publicly available with the collaboration release.

\appendix
\setcounter{equation}{0}
\renewcommand{\theequation}{A.\arabic{equation}}

\section{Origin of Eqs.~(\ref{eq:fried1})--(\ref{eq:fried2}) and of the scaling $\Phi\propto a^{-3}$} \label{app:derivation}

This appendix collects, for self-containedness, the EC pedigree of the Friedmann pair \eqref{eq:fried1}--\eqref{eq:fried2}, its relation to other torsion cosmologies, and the microphysical reading of the adiabatic scaling~\eqref{eq:phia}.

\subsection*{A.1 \ Torsion in a Friedmann universe}

In EC gravity the connection is metric-compatible but carries torsion, $T^{\lambda}{}_{\mu\nu}=\Gamma^{\lambda}{}_{\mu\nu}-\Gamma^{\lambda}{}_{\nu\mu}$, and the action is the Einstein--Hilbert one built on the full connection, plus matter. Independent variation of the metric (or tetrad) and the connection yields an Einstein equation with torsion corrections together with the Cartan equation, which is \emph{algebraic}: torsion is fixed pointwise by the spin tensor of matter and vanishes with it~\cite{Hehl1976}. Torsion does not propagate; its cosmological effects ride on the matter that sources it. Gravitational waves in this setting remain luminal, with torsion affecting polarizations and amplitudes at levels far below current sensitivity, so that compact-binary mergers remain standard sirens~\cite{Barrientos2019,Elizalde2023}; the late universe is, in this sense too, remarkably indifferent to torsion.

Spatial homogeneity and isotropy restrict the torsion tensor drastically. Two independent modes survive, each described by a single function of time: a \emph{vectorial} mode, in which torsion points along the fundamental four-velocity (this is the $\phi(t)$ of~\cite{KTBI2019}), and an \emph{axial} (totally antisymmetric) mode, $T_{ijk}\propto f(t)\,\varepsilon_{ijk}$, whose dual is a timelike pseudovector. The two behave very differently in the field equations. The vectorial mode couples linearly to the expansion: the Friedmann and continuity equations of~\cite{KTBI2019} contain $\phi H$ and $\phi\rho$ cross terms, which is what makes ``steady'' scenarios $\phi\propto H$ possible~\cite{KTBI2019,Liu2025}. The axial mode cannot do this: a parity-odd vector has no business appearing linearly in a parity-even scalar equation, so it enters the Friedmann equations only through its square. It is also the mode sourced by the totally antisymmetric spin tensor of Dirac fields~\cite{Hehl1976}, which is why it is the natural carrier of the ``spin of matter'' in cosmology.

\subsection*{A.2 \ Spin fluid and the effective Friedmann pair}

The classic macroscopic source is the Weyssenhoff spin fluid: particles carrying intrinsic spin $s_{\mu\nu}$, randomly oriented so that $\langle s_{\mu\nu}\rangle=0$ while the mean square spin density $\sigma^{2}\propto\langle s_{\mu\nu}s^{\mu\nu}\rangle$ survives. Substituting the Cartan equation into the Einstein equation and averaging, the linear spin terms drop out and the quadratic ones assemble into an effective perfect fluid (see, e.g.,~\cite{Brechet2008,Poplawski2012}),
\begin{equation}
\rho_{\rm eff}=\rho-\frac{\kappa\,\sigma^{2}}{4},
\qquad
p_{\rm eff}=p-\frac{\kappa\,\sigma^{2}}{4},\label{eq:appeff}
\end{equation}
with $\kappa=8\pi G=1$ in our units. The standard FLRW equations for $(\rho_{\rm eff},p_{\rm eff})$ then read
\begin{equation}
3H^{2}=\rho-\frac{\sigma^{2}}{4},
\qquad
\dot H+H^{2}=-\frac{1}{6}\left(\rho+3p-\sigma^{2}\right), \label{eq:appfried}
\end{equation}
which are precisely Eqs.~\eqref{eq:fried1}--\eqref{eq:fried2} under the identification
\begin{equation}
\Phi^{2}=\frac{\sigma^{2}}{12} . \label{eq:appmap}
\end{equation}
Normalizations of $\sigma^{2}$ vary across the literature, and they only move the numerical factor in \eqref{eq:appmap}; the structure (negative, quadratic in the spin density, stiff) is convention-independent. For elementary fermions the route is different in detail and identical in outcome: integrating out the algebraic axial torsion generates the well-known four-fermion contact interaction, whose FLRW average is again a negative quadratic correction~\cite{Hehl1976,Poplawski2012}. In either reading, Eqs.~\eqref{eq:fried1}--\eqref{eq:fried2}, which we took operationally from~\cite{YunLee2026}, are the standard spin-sourced EC cosmology: $\Phi$ measures the axial torsion amplitude, equivalently the spin density of its source.

\subsection*{A.3 \ The adiabatic scaling as kinematics}

For a comoving fluid, particle-number conservation, $\nabla_{\mu}(n\,u^{\mu})=0$, gives $n\propto a^{-3}$. The spin density is $\sigma=\langle s\rangle\,n$, with $\langle s\rangle$ the mean spin per particle set by microphysics ($\hbar/2$ times an alignment factor). If $\langle s\rangle$ is constant, then $\sigma\propto n\propto a^{-3}$ and, through \eqref{eq:appmap}, $\Phi\propto a^{-3}$: Eq.~\eqref{eq:phia} is the kinematics of dilution. Matter conservation and $\dot\Phi+3H\Phi=0$ are then two faces of the same statement, which is why they solve the exchange identity \eqref{eq:bianchi} simultaneously, with both sides vanishing. To make $\Phi$ decay more slowly than $a^{-3}$, the spin per particle $\langle s\rangle$ must grow with time: a progressive alignment of spins, which is precisely the kind of interacting scenario discussed in Sec.~\ref{sec:discussion} and explored for the dark matter spin tensor in~\cite{ILV2020,Elizalde2023}.

\subsection*{A.4 \ Notation map}

Our $\Phi$ is that of~\cite{YunLee2026} (their Eqs.~(3.28)--(3.29), arXiv v1), up to the sign convention for the density parameter discussed in Sec.~\ref{sec:nogo}. The $\phi$ of~\cite{KTBI2019} is the vectorial mode and belongs to a different family: it couples linearly to $H$, and the parametrization $\phi=-\alpha H/2$ constrained in~\cite{Liu2025} modifies the matter scaling itself ($\rho\propto a^{-3+\alpha}$), with no analogue of the bounce structure studied here. The Weyssenhoff spin density $\sigma$ maps onto $\Phi$ through \eqref{eq:appmap}.


\begin{thebibliography}{99}

\bibitem{DESI-DR2}
DESI Collaboration, M.~Abdul-Karim \emph{et al.}, ``DESI DR2 Results II: Measurements of Baryon Acoustic Oscillations and Cosmological Constraints,'' arXiv:2503.14738 [astro-ph.CO], Phys.\ Rev.\ D (2025).

\bibitem{Hehl1976}
F.~W.~Hehl, P.~von der Heyde, G.~D.~Kerlick and J.~M.~Nester, ``General relativity with spin and torsion: Foundations and prospects,'' Rev.\ Mod.\ Phys.\ \textbf{48} (1976) 393.

\bibitem{Trautman1973}
A.~Trautman, ``Spin and torsion may avert gravitational singularities,'' Nature Phys.\ Sci.\ \textbf{242} (1973) 7--8.

\bibitem{Poplawski2012}
N.~J.~Pop\l awski, ``Nonsingular, big-bounce cosmology from spinor-torsion coupling,'' Phys.\ Rev.\ D \textbf{85} (2012) 107502, arXiv:1111.4595 [gr-qc].

\bibitem{Brechet2008}
S.~D.~Brechet, M.~P.~Hobson and A.~N.~Lasenby, ``Classical big-bounce cosmology: dynamical analysis of a homogeneous and irrotational Weyssenhoff fluid,'' Class.\ Quantum Grav.\ \textbf{25} (2008) 245016, arXiv:0807.2523 [gr-qc].

\bibitem{KTBI2019}
D.~Kranas, C.~G.~Tsagas, J.~D.~Barrow and D.~Iosifidis, ``Friedmann-like universes with torsion,'' Eur.\ Phys.\ J.\ C \textbf{79} (2019) 341, arXiv:1809.10064 [gr-qc].

\bibitem{TS2011}
A.~Tilquin and T.~Sch\"ucker, ``Torsion, an alternative to dark matter?,'' Gen.\ Rel.\ Grav.\ \textbf{43}  2965–2978 (2011),  arXiv:1104.0160 [astro-ph.CO].

\bibitem{Pereira2022}
S.~H.~Pereira, A.~M.~Vicente, J.~F.~Jesus and R.~F.~L.~Holanda, ``Dark matter from torsion in Friedmann cosmology,'' Eur.\ Phys.\ J.\ C \textbf{82} (2022) 356, arXiv:2202.01807 [gr-qc].

\bibitem{Liu2025}
T.~Liu, X.~Li, T.~Xu, M.~Biesiada and J.~Wang, ``Torsion cosmology in the light of DESI, supernovae and CMB observational constraints,'' Eur.\ Phys.\ J.\ C \textbf{85}:1351 (2025), arXiv:2507.04265 [astro-ph.CO].

\bibitem{SNY2008}
K.-F.~Shie, J.~M.~Nester and H.-J.~Yo, ``Torsion cosmology and the accelerating universe,'' Phys.\ Rev.\ D \textbf{78} (2008) 023522, arXiv:0805.3834 [gr-qc].

\bibitem{YunLee2026}
Y.~Yun and J.~Lee, ``Holographic Dark Energy with Hubble Radius as an Infrared Cutoff in Einstein-Cartan Gravity,'' arXiv:2605.22143 [gr-qc].

\bibitem{YunLee2024}
Y.~Yun and J.~Lee, ``Holographic Dark Energy with Torsion,'' arXiv:2407.16992 [gr-qc].

\bibitem{LiChen2023}
S.~Li and Y.~Chen, ``Reconstructing Torsion Cosmology from Interacting Holographic Dark Energy Model,'' Universe \textbf{9} (2023) 100, doi:10.3390/universe9020100.

\bibitem{Li2004}
M.~Li, ``A model of holographic dark energy,'' Phys.\ Lett.\ B \textbf{603} (2004) 1, arXiv:hep-th/0403127.

\bibitem{Hsu2004}
S.~D.~H.~Hsu, ``Entropy bounds and dark energy,'' Phys.\ Lett.\ B \textbf{594} (2004) 13, arXiv:hep-th/0403052.

\bibitem{GO2008}
L.~N.~Granda and A.~Oliveros, ``Infrared cut-off proposal for the holographic density,'' Phys.\ Lett.\ B \textbf{669} (2008) 275, arXiv:0810.3149 [gr-qc].

\bibitem{Carniani2024}
S.~Carniani \emph{et al.}, ``Spectroscopic confirmation of two luminous galaxies at a redshift of 14,'' Nature \textbf{633} (2024) 318, arXiv:2405.18485 [astro-ph.GA].

\bibitem{Planck2018}
Planck Collaboration, N.~Aghanim \emph{et al.}, ``Planck 2018 results. VI. Cosmological parameters,'' Astron.\ Astrophys.\ \textbf{641} (2020) A6, arXiv:1807.06209 [astro-ph.CO].

\bibitem{Fields2020}
B.~D.~Fields, K.~A.~Olive, T.-H.~Yeh and C.~Young, ``Big-Bang Nucleosynthesis after Planck,'' JCAP \textbf{03} (2020) 010, arXiv:1912.01132 [astro-ph.CO].

\bibitem{Lepe2022}
M.~Cruz, S.~Lepe and G.~Soto, ``Phantom cosmologies from QCD ghost dark energy,'' Phys.\ Rev.\ D \textbf{106} (2022) 103508, arXiv:2209.04584 [gr-qc].

\bibitem{Lepe2023}
M.~Cruz and S.~Lepe, ``Thermodynamics of a transient phantom scenario,'' Phys.\ Dark Univ.\ \textbf{42} (2023) 101367, arXiv:2304.02735 [gr-qc].

\bibitem{Lepe2026}
M.~Cruz, S.~Lepe and J.~Saavedra, ``The holographic origin of future singularities and the role of spatial curvature in cosmic expansion,'' Fortschr.\ Phys.\ \textbf{74} (2026) e70087, arXiv:2512.07791 [gr-qc].

\bibitem{ILV2020}
F.~Izaurieta, S.~Lepe and O.~Valdivia, ``The Spin Tensor of Dark Matter and the Hubble Parameter Tension,'' Phys.\ Dark Univ.\ \textbf{30} (2020) 100662, arXiv:2004.13163 [gr-qc].

\bibitem{Barrientos2019}
J.~Barrientos, F.~Cordonier-Tello, C.~Corral, F.~Izaurieta, P.~Medina, E.~Rodr\'iguez and O.~Valdivia, ``Luminal propagation of gravitational waves in scalar-tensor theories: The case for torsion,'' Phys.\ Rev.\ D \textbf{100} (2019) 124039, arXiv:1910.00148 [gr-qc].

\bibitem{Elizalde2023}
E.~Elizalde, F.~Izaurieta, C.~Riveros, G.~Salgado and O.~Valdivia, ``Gravitational waves in Einstein--Cartan theory: On the effects of dark matter spin tensor,'' Phys.\ Dark Univ.\ \textbf{40} (2023) 101197, arXiv:2204.00090 [gr-qc].

\bibitem{emcee}
D.~Foreman-Mackey, D.~W.~Hogg, D.~Lang and J.~Goodman, ``emcee: The MCMC Hammer,'' Publ.\ Astron.\ Soc.\ Pac.\ \textbf{125} (2013) 306, arXiv:1202.3665 [astro-ph.IM].

\end{thebibliography}
\end{document}